# Radiative heat transfer in low-symmetry Bravais crystal

Cheng–Long Zhou,[1,2] Gao-Ming Tang,[3] Yong Zhang,[1,2*] Mauro Antezza,[4,5] and Hong–Liang Yi[1,2†]

[1]*School of Energy Science and Engineering, Harbin Institute of Technology, Harbin 150001, People's Republic of China*
[2]*Key Laboratory of Aerospace Thermophysics, Ministry of Industry and Information Technology, Harbin 150001, People's Republic of China*
[3]*Graduate School of China Academy of Engineering Physics, Beijing 100193, China*
[4]*Laboratoire Charles Coulomb (L2C), UMR 5221 CNRS-Université de Montpellier, F- 34095 Montpellier, France*
[5]*Institut Universitaire de France, 1 rue Descartes, F-75231 Paris, France*

**ABSTRACT:** Over the last few years, broken symmetry within crystals has attracted extensive attention since it can improve the control of light propagation. In particular, low-symmetry Bravais crystal can support shear polaritons which has great potential in thermal photonics. In this work, we report a twist-induced near-field thermal control system based on the low-symmetry Bravais crystal medium ($\beta$-$Ga_2O_3$). The near-field thermal radiation (NFTR) between such crystal slabs is nearly four orders of magnitude larger than the blackbody limit, exceeding the NFTR from other traditional dielectric materials. Moreover, we show that this crystal can serve as an excellent platform for twist-induced near-field thermal control. Due to the intrinsic shear effect, the twist-induced modulation supported by low-symmetry Bravais crystal exceeds that by high-symmetry crystal. We further clarify how the shear effect affects the twist-induced thermal-radiation modulation supported by hyperbolic and elliptical polaritons and show that the shear effect significantly enhances the twist-induced thermal control induced by the elliptical polariton mode. These results open new directions for thermal-radiation control in low-symmetry materials, including geological minerals, common oxides, and organic crystals.

**KEYWORDS**: Near-field radiative modulation, Low-symmetry Bravais crystal, Shear polaritons.

## 1. INTRODUCTION

In contrast with classical thermal radiation [1-3], evanescent waves coupling between two bodies with subwavelength separation can assist photons to tunnel through a vacuum gap (i.e., photon tunneling), resulting in a radiative heat flux that can exceed the blackbody limit by several orders of magnitude [4, 5]. Over the last two decades, this is the so-called near-field thermal radiation (NFTR) has inspired significant research interest because of its fundamental scientific relevance [6–8], and it is particularly important for technological applications, such as thermal logic circuitry [9–13], high-efficiency energy conversion [14–18], photon transformation [19], and magnetic recording [20]. For these novel applications, it is important to actively control NFTR. Several active control strategies have been proposed and are based on modifying the dielectric function of materials to change their dispersion relation and surface polaritons; these include the use of insulator-metal transition materials [21, 22], applying an electric field to ferroelectric materials [23], the use of an external magnetic field to magneto-optical materials [24-26], the use of drift currents [27, 28], and the regulation of the chemical potential of photons [29–31]. Another active control strategy is to use the rotational degrees of freedom. In analogy to the twistronics concept in flat-band superconductivity [32], polariton topological transitions [33, 34], and Moiré excitons [35], this control strategy is also called the twisting method. To date, most proposals for implementing the twisting method rely strongly on processing technology for nanometer-sized metamaterials [36–39] or external field assistance (electric field, magnetic field, tension, etc.) [40–42]. Creating a convenient and efficient platform for controlling heat flux remains an open challenge.

---

[*] Email: yong_zhang@hit.edu.cn
[†] Email: yihongliang@hit.edu.cn

Recently, low-symmetry crystals have drawn widespread attention due to their unique physical properties, such as nonlinear frequency conversion [43], structural phase transitions [44], ferroelectrics [45], quantum memories [46], and slow light [47]. In particular, low-symmetry Bravais crystals exhibit nonorthogonal principal crystal axes, resulting in the nontrivial relative orientation (neither parallel nor orthogonal) of several optical transitions [48]. Consequently, their dielectric permittivity tensor has major polarizability directions that depend strongly on the frequency, with off-diagonal terms that cannot be completely removed through coordinate rotation. The tensor also has shear terms analogous to the case of viscous flow [49]. This shear phenomenon in the dielectric response induces extreme anisotropic propagation of surface waves, giving rise to a new polariton class of "shear polaritons". All of these factors make low-symmetry Bravais crystals an exciting platform with strong anisotropic polaritons, which opens avenues for the manipulation of polariton propagation, light-matter interaction, and light emission. However, fundamental studies of such low-symmetry Bravais crystals remain scarce, especially in the field of thermal radiation, which hinders the further exploration and practical application of these materials.

To address this issue, we investigate beta gallium oxide ($\beta$-Ga$_2$O$_3$) as an example of a low-symmetry Bravais crystal and reveal the mechanisms of radiative heat transport in $\beta$-Ga$_2$O$_3$ by applying the fluctuation-dissipation theorem. Moreover, we explore the twist-induced control of thermal radiation in the low-symmetry Bravais crystal system. By introducing a scaling factor to describe the magnitude of the off-diagonal component, we tune the shear phenomenon to modulate the photon tunneling features in the search for guidance in radiation modulation. The results are summarized in the conclusion.

## 2. SYSTEM UNDER STUDY AND THEORETICAL APPROACH

We consider the near-field thermal radiation model consisting of two parallel $\beta$-Ga$_2$O$_3$ slabs with a vacuum gap $d$ [see Fig. 1(a)]. The temperature of the top and bottom $\beta$-Ga$_2$O$_3$ slabs is $T+\Delta T$ and $T$, respectively, with $T$ = 300 K. The thickness of the $\beta$-Ga$_2$O$_3$ slabs is 100 μm. To gain insight into the crystalline structure of $\beta$-Ga$_2$O$_3$, we present in Fig. 1(c) the crystal structure of the monoclinic plane $a$-$c$. Due to the monoclinic angle ($\beta$ = 103.7°) [50–52], axes $a$ and $c$ are not perpendicular to each other, as shown in Fig. 1(c). A Cartesian coordinate system $x$-$y$-$z$ ($x'$-$y'$-$z$) is defined for the bottom (top) slab, and the $x$ ($x'$) and $z$ axes are in the direction of axes $a$ and $b$ of $\beta$-Ga$_2$O$_3$, respectively. The twist angle $\theta$ is the angle between the $x$ and $x'$ axes [see Fig. 1(b)].

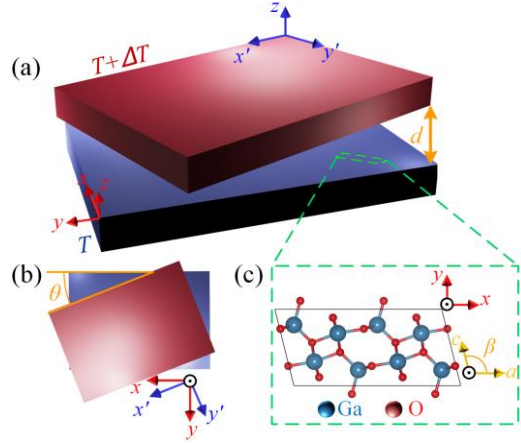

**Fig. 1.** (a) Setup for near-field radiative modulator composed of two beta-phase $Ga_2O_3$ with vacuum gap $d$ and twist angle $\theta$. (b) Schematic view from the top of the twist-induced thermal radiation modulator. The surface of the $\beta$-$Ga_2O_3$ crystal is the monoclinic (010) plane ($x$-$y$ plane). (c) Monoclinic crystal structure of $\beta$-$Ga_2O_3$ in $a$-$c$ plane ($x$-$y$ plane).

In this work, we focus on analyzing the heat transfer coefficient (HTC) $H$ which is the radiative heat conductance per unit area, [53]

$$H(T,d)=\int_0^\infty h(\omega)d\omega = \frac{1}{8}\int_0^\infty \frac{\partial \Theta(\omega,T)}{\partial T}d\omega \int_0^\infty \int_0^\infty \xi(\omega,k_x,k_y)dk_x dk_y, \quad (1)$$

where $\Theta(\omega,T) = \hbar\omega/[\exp(\hbar\omega/k_B T)-1]$ is the mean photon energy of a Planck oscillator at angular frequency $\omega$ and temperature $T$, and $k_x$ and $k_y$ are the surface wavevectors along the $x$ and $y$ axes, respectively. When the surface parallel wavevector $k=(k_x^2+k_y^2)^{1/2}$ is greater than the wavevector in vacuum, $k_0 = \omega/c_0$, the electromagnetic wave excited by thermal energy takes the form of an evanescent wave [54]; otherwise, it is propagating wave [55]. $\xi(\omega,k_x,k_y)$ is the photonic tunneling coefficient (PTC) of thermal photons and is expressed as [56]

$$\xi(\omega,k_x,k_y)=\begin{cases} \text{Tr}[(\mathbf{I}-\mathbf{R}_2^*\mathbf{R}_2-\mathbf{T}_2^*\mathbf{T}_2)\mathbf{D}(\mathbf{I}-\mathbf{R}_1\mathbf{R}_1^*-\mathbf{T}_1^*\mathbf{T}_1)\mathbf{D}^*], & k<k_0 \\ \text{Tr}[(\mathbf{R}_2^*-\mathbf{R}_2)\mathbf{D}(\mathbf{R}_1-\mathbf{R}_1^*)\mathbf{D}^*]e^{-2|k_{z0}|d}, & k>k_0. \end{cases} \quad (2)$$

The identity matrix is denoted $\mathbf{I}$, subscripts 1 and 2 indicate the emitter (bottom slab) and receiver (top slab), respectively, and $k_{z0}=(k_0^2-k^2)^{1/2}$ is the tangential wavevector along the $z$ direction in vacuum. The reflection coefficient matrix $\mathbf{R}_n$ and the transmission coefficient matrix $\mathbf{T}_n$ at the interface between air and $\beta$-$Ga_2O_3$ slab with $n=1, 2$ take the forms [57, 58]:

$$\mathbf{R}_n = \begin{bmatrix} r_n^{ss} & r_n^{sp} \\ r_n^{ps} & r_n^{pp} \end{bmatrix}, \quad \mathbf{T}_n = \begin{bmatrix} t_n^{ss} & t_n^{sp} \\ t_n^{ps} & t_n^{pp} \end{bmatrix} \quad (3)$$

and can be calculated by the transfer matrix method [59]. The 2×2 matrix $\mathbf{D}$ is defined as $\mathbf{D} \equiv (\mathbf{I}-\mathbf{R}_1\mathbf{R}_2 e^{2ik_{z0}d})^{-1}$, which describes the usual Fabry–Perot-like denominator resulting from multiple scattering between emitter and receiver [60, 61]. In general, the frequency-dependent permittivity of $\beta$-$Ga_2O_3$ can be expressed as a third-rank tensor with identical off-diagonal elements [48]:

$$\bar{\bar{\varepsilon}} = \begin{bmatrix} \varepsilon_{xx} & \varepsilon_{xy} & 0 \\ \varepsilon_{xy} & \varepsilon_{yy} & 0 \\ 0 & 0 & \varepsilon_{zz} \end{bmatrix}. \tag{4}$$

The $\beta$-Ga$_2$O$_3$ monoclinic permittivity tensor is composed of the high-frequency contributions, the dipole charge resonances, and the free charge carrier contributions [50]:

$$\varepsilon_{xx} = \varepsilon_{\infty,xx} + \sum_{j=1}^{8} \rho_{j,B_u} \cos^2 \alpha_j + \vartheta_{FCC,x}, \tag{5a}$$

$$\varepsilon_{yy} = \varepsilon_{\infty,yy} + \sum_{j=1}^{8} \rho_{j,B_u} \sin^2 \alpha_j + \vartheta_{FCC,y}, \tag{5b}$$

$$\varepsilon_{zz} = \varepsilon_{\infty,zz} + \sum_{j=1}^{8} \rho_{j,A_u} + \vartheta_{FCC,z}, \tag{5c}$$

$$\varepsilon_{xy} = \varepsilon_{\infty,xy} + \sum_{j=1}^{8} \rho_{j,B_u} \cos \alpha_j \sin \alpha_j, \tag{5d}$$

where $\varepsilon_\infty$ is the permittivity at high frequency for different directions, the $A_u$ and $B_u$ modes are long-wavelength active (infrared and far-infrared); the $A_u$ modes are polarized along axis $b$ only and the $B_u$ modes are polarized within the $a$-$c$ plane. $\alpha_j$ is the angle between the dipole oscillation axis of $B_u$ mode $j$ and the axis $a$ of the $\beta$-Ga$_2$O$_3$ monoclinic crystal. The Lorentzian-broadened oscillator function $\rho_j$ represents the energy-dependent contribution to the long-wavelength polarization response of an uncoupled electric dipole charge oscillation of $B_u$ mode $j$ ($A_u$ mode). The Drude model function $\vartheta_{FCC}$ gives the energy-dependent contribution to the long-wavelength polarization response of free charge carriers for different directions. The detailed parametrization of the dielectric function of $\beta$-Ga$_2$O$_3$ monoclinic crystal is available in Ref. [50]. The $\beta$-Ga$_2$O$_3$ monoclinic crystals are Sn-doped to produce a free charge carrier density of $n_i = 4 \times 10^{17}$ cm$^{-3}$, which is reasonable in $\beta$-Ga$_2$O$_3$ samples [62]. The diagonal permittivity elements of a $\beta$-Ga$_2$O$_3$ monoclinic crystal are plotted in Fig. 2(a). In biaxial crystals, the sign of the real part of the diagonal permittivity elements determines the topology of the polariton mode. All diagonal permittivity elements have a negative real part, producing an elliptical polariton mode. When the diagonal permittivity elements change sign, a polariton mode of hyperbolic topology is produced in which either the real part of one element is negative and the other two are positive (type-I hyperbolic mode) or two are negative and one is positive (type-II hyperbolic mode). However, for the monoclinic crystal, the effect of the dielectric response of off-diagonal elements on the polariton nature is non-negligible. This strong dielectric response of off-diagonal elements typically prevents the electromagnetic propagation angle from being aligned with the principal axes, leading to a non-trivial polariton response with highly directional modes and giving rise to a remarkable shearing effect of polariton modes in monoclinic crystals [48].

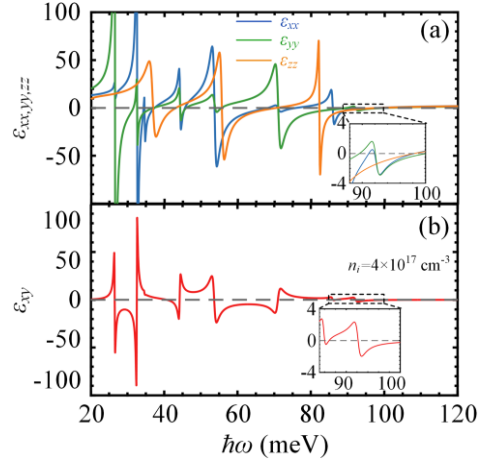

**Fig. 2.** (a) Real part of diagonal permittivity element $\varepsilon_{xx,yy,zz}$. (b) Real part of off-diagonal permittivity element $\varepsilon_{xy}$. The free charge carrier density $n_i$ is fixed at $4\times10^{17}$ cm$^{-3}$.

## 3. RESULTS AND DISCUSSIONS

We first investigate the NFTR of this low-symmetry Bravais crystal system without twist. Figure 3(a) shows the HTC as a function of the vacuum gap $d$. To visualize the near-field enhancement for this material, Fig. 3(a) also shows the HTC of a blackbody at room temperature given by $H_{BB} = 4\sigma_{SB}T^3 = 6.12$ W m$^{-2}$ K$^{-1}$, where $\sigma_{SB}$ is the Stefan–Boltzmann constant. The HTC of the $\beta$-Ga$_2$O$_3$ system is colossally enhanced in the near-field region due to the evanescent contribution, which is several orders of magnitude greater than the blackbody limit. In particular, the results show that the HTC of such crystals is 8264 times the blackbody limit at $d = 10$ nm. As shown in Fig. 3(a), when the gap is less than 400 nm, the HTC of this crystal approaches a $d^{-2}$ dependence (plotted by the blue curve). This dependence is consistent with the approximate expression in Ref. [63], in which the NFTR for any bulk materials tends to follow the unique asymptotic law of $d^{-2}$ in the deep near field.

The HTC of the $\beta$-Ga$_2$O$_3$ system decreases notably to 1.42$h_{BB}$ as the gap approaches 1 μm in Fig. 3(a) because the large gap severely limits the evanescent contribution to thermal radiation. A spectral analysis allows us to understand the mechanism for the NFTR. Figure 3(b) shows that the spectral HTC for this low-symmetry Bravais crystal at $d = 10$ nm has a prominent feature with multiple peaks. This is also well predicted by Fig. 2, which shows that this material has multiple resonance modes in the mid- and far-infrared region, complicating the polariton behavior in this low-symmetry Bravais crystal. Thanks to this multi-peak feature, the spectral HTC of this low-symmetry Bravais crystal can cover a broad frequency region, which is particularly important for obtaining a strong NFTR. Moreover, Fig. 3(b) shows that the spectral HTC features three main peaks, playing an important role in the NFTR. Combined with Fig. 2, we see that the three peaks are in the hyperbolic-mode region from 55 to 69 meV/$\hbar$, the hyperbolic-mode region from 89 to 91 meV/$\hbar$, and the elliptical-mode region from 92 to 98 meV/$\hbar$, respectively. In the subsequent analysis, for the sake of discussion, we focus our attention entirely on these three frequency ranges.

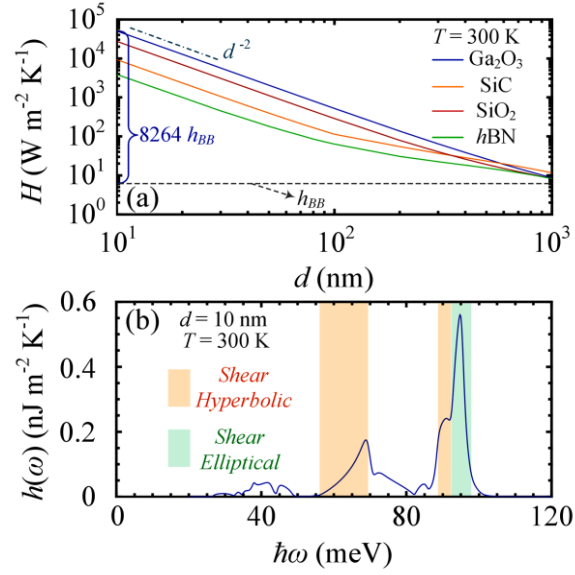

**Fig. 3.** (a) Heat-transfer coefficient $H$ of $\beta$-Ga$_2$O$_3$ system as a function of the vacuum gap $d$. (b) Spectral HTC $h(\omega)$ as a function of energy for a vacuum gap of 10 nm. The gray dotted line in panel (a) is the HTC for the blackbody. The hyperbolic and the elliptical regimes in panel (b) are represented by the yellow and green areas, respectively. The temperature is fixed at 300 K.

Additionally, to visually evaluate the NFTR of this low-symmetry Bravais crystal, the HTCs of other types of terminal structures are also shown in Fig. 3(a) [(I) SiC slab, (II) SiO$_2$ slab, and (III) $h$BN slab]. In this section, the thicknesses and temperatures of all samples are fixed at 100 μm and 300 K. The optical properties of SiC, SiO$_2$, and $h$BN are modeled as per Refs. [64–66]. Note also that, as representative platforms with robust polariton modes, these materials are widely used in the research of NFTR [67–70]. As shown in Fig. 3(a), the HTCs of the SiC and SiO$_2$ slabs are 9.29 and 27.37 kW m$^{-2}$K$^{-1}$, respectively, for $d$ = 10 nm. Additionally, the HTC produced by the $h$BN slab only is 3.89 kW m$^{-2}$K$^{-1}$ at the same vacuum gap. Interestingly, Fig. 3(a) shows that the $\beta$-Ga$_2$O$_3$ slab can yield a greater HTC than these typical polariton materials, especially for a small vacuum gap. These results show that this low-symmetry Bravais crystal produces $H$ = 50.61 kW m$^{-2}$K$^{-1}$ for $d$ = 10 nm, which is 13.01 times greater than the HTC of an individual $h$BN slab. In other words, this low-symmetry Bravais crystal offers a larger potential for noncontact heat dissipation for nanoscale circuits or other devices.

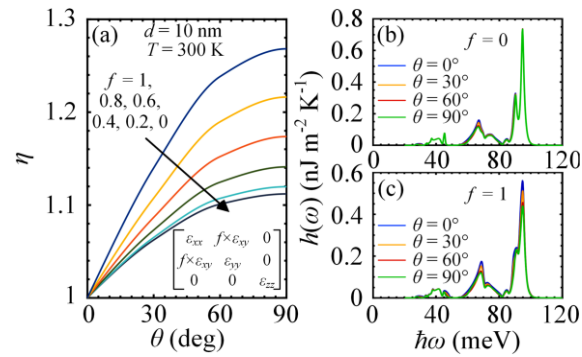

**Fig. 4.** (a) Twist-induced modulated coefficient $\eta$ at different scaling factors for the magnitude of the off-diagonal component. (b) Spectral HTC for different twist angles in shear-free biaxial material ($f$ = 0). (c) Spectral HTC for different twist angles in low-symmetry crystal ($f$ = 1). The vacuum gap is 10 nm and the temperature is 300 K.

To determine whether the low-symmetry Bravais crystals can provide a highly tunable platform for near-field radiative heat transfer, we now discuss the twist-induced near-field thermal modulation induced by intrinsic shear of the surface modes in $β$-Ga$_2$O$_3$. The vacuum gap is 10 nm. To visually analyze the role of the shear phenomenon in thermal modulation, we introduce a scaling factor $f$ to describe the magnitude of the off-diagonal component, indicated as $fε_{xy}$. The scaling factor shows the intensity of symmetry breaking for propagating polaritons. When we remove the off-diagonal component ($f$ = 0), $β$-Ga$_2$O$_3$ becomes a shear-free biaxial material in which mode propagation is symmetric with respect to the crystal axes, similar to $α$-MoO$_3$. As the scaling factor for the off-diagonal component returns to its natural value ($f$ = 1), the propagating polaritons become increasingly skewed from the crystal axes. A similar analysis has also been applied to shear hyperbolic polaritons [48]. This factor provides evidence that intrinsic shear in low-symmetry Bravais crystals can improve the twist-induced near-field thermal modulation, as shown in Fig. 4(a). When there is no shear phenomenon ($f$ = 0), as the twist angle increases from 0º to 90º, the twist-induced modulated coefficient $η$ reaches 1.11. Note that the twist-induced modulated coefficient $η$ is normalized by $h(0°)/h(θ)$ for a given scaling factor.

Figure 4(a) shows that the twist-induced modulated coefficient is enhanced upon increasing the scaling factors for the magnitude of the off-diagonal component. As the scaling factor for the off-diagonal component returns to its natural value, the maximal twist-induced modulated coefficient increases to 1.27 for $θ$ = 90°, which is 1.15 times greater than that in the shear-free scenario. Given that a greater scaling factor generates a stronger shear effect, we predict that intrinsic shear in low-symmetry Bravais crystals can modulate the twist-induced near-field thermal radiation. A spectral analysis allows us to understand the mechanism for the enhanced twist-induced near-field thermal control induced by the intrinsic shear. Figures 4(b) and 4(c) show the spectral HTCs of the shear-free biaxial material ($f$ = 0) and of $β$-Ga$_2$O$_3$ ($f$ = 1), respectively. Note that, when the scaling factor increases from zero to one, the spectral HTC is attenuated at all frequencies, as shown by Figs. 4(b) and 4(c). Additionally, note that introducing off-diagonal components enhances the twist-induced near-field thermal control in the elliptical region (from 92 to 98 meV/$ℏ$). Consider the spectral HTC at $ω$ = 94.5 meV/$ℏ$ as an example. For the shear-free biaxial material, at the given frequency, the change in the spectral HTC upon increasing the twist angle is difficult to detect [Fig. 4(b)]. This also means that, at the given frequency, for zero off-diagonal components for this material, the surface polaritons have a very weak anisotropy and thus cannot exert a strong twist-induced near-field thermal control. Fortunately, the introduction of an off-diagonal component improves the twist-induced near-field thermal control due to support by the elliptical polariton mode, as seen in Fig. 4(c). The spectral HTC decreases from 0.56 nJ m$^{-2}$ K$^{-1}$ at $θ$ = 0° to 0.42 nJ m$^{-2}$ K$^{-1}$ at $θ$ = 90° [Fig. 4(c)].

However, this enhancement of twist-induced thermal control does not appear in the frequency region dominated by hyperbolic polaritons. On the contrary, the introduction of the off-diagonal component weakens the twist-induced thermal control supported by the hyperbolic polaritons. Consider the spectral HTC at $\omega = 67$ meV/$\hbar$ as an example. For the shear-free biaxial material, the spectral HTC decreases from 0.16 nJ m$^{-2}$K$^{-1}$ at $\theta = 0°$ to 0.12 nJ m$^{-2}$K$^{-1}$ at $\theta = 90°$ [Fig. 4(b)]. As the scaling factor for the off-diagonal component returns to its natural value, the spectral HTC only decreases from 0.14 nJ m$^{-2}$K$^{-1}$ at $\theta = 0°$ to 0.11 nJ m$^{-2}$K$^{-1}$ at $\theta = 90°$ [Fig. 4(c)].

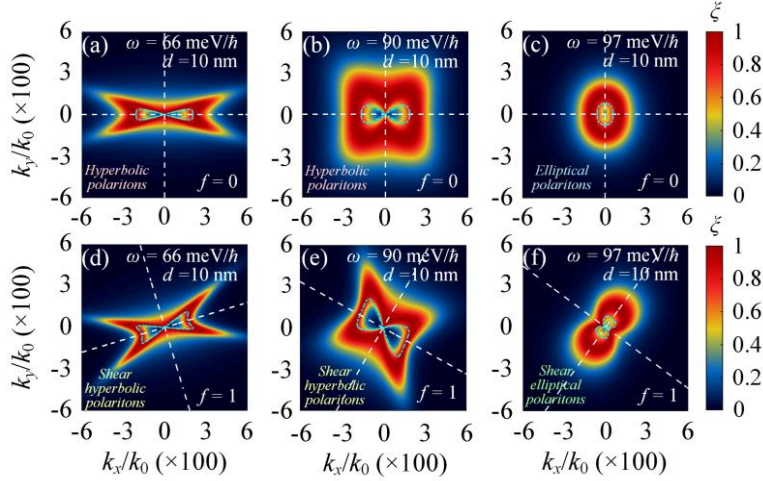

**Fig. 5.** PTC at a frequency of (a) 66 meV/$\hbar$, (b) 90 meV/$\hbar$, and (c) 97 meV/$\hbar$ in symmetry crystal systems ($f = 0$) and PTC at a frequency of (d) 66 meV/$\hbar$, (e) 90 meV/$\hbar$, and (f) 97 meV/$\hbar$ in low-symmetry crystal systems ($f = 1$). The blue dashed lines represent the polaritons resonance dispersion curves of the system. The white dashed lines indicate the optical axis.

The mechanisms by which shearing affects the twist-induced near-field thermal control can be explored by photon tunneling between the two $\beta$-Ga$_2$O$_3$ slabs (i.e., the PTC). We first demonstrate how off-diagonal components (i.e., the shearing effect) influence the hyperbolic polariton resonance and the elliptical polariton resonance (see Fig. 5) when the system is in the parallel case ($\theta = 0°$). Here, we show the specific cases at 66 meV/$\hbar$ and 90 meV/$\hbar$ to reveal how shear affects the hyperbolic polariton resonance. As shown in Fig. 2, at 66 meV/$\hbar$ and 90 meV/$\hbar$, the polariton nature of this low-symmetry Bravais crystal is unquestionably located in the region of the hyperbolic-mode polaritons. Figure 5(a) shows the in-plane PTC for the shear-free biaxial material system at $\omega = 66$ meV/$\hbar$. At this frequency, the diagonal permittivity elements are $\varepsilon_{xx} = -1.92 + 0.77i$, $\varepsilon_{yy} = 14.51 + 2.35i$, and $\varepsilon_{zz} = -3.01 + 0.79i$. A hyperbolic contour of the PRC is also shown in Fig. 5(a). For the shear-free biaxial material, these surface waves reveal an in-plane hyperbolic feature and are thus referred to as "hyperbolic surface" or "hyperbolic Dyakonov" polaritons [71], in which the hyperbolic polariton response supported by a shear-free biaxial material is symmetric with respect to the crystal axes [Fig. 5(a)].

To confirm the dominant role of hyperbolic polaritons in the photon tunneling of shear-free biaxial materials, we show in Fig. 5(a) the in-plane polariton resonance dispersion at different frequencies. Since the contributions to NFTR of dielectric materials are dominated by TM waves [72, 73], the dispersion relation is obtained from $1 - r^{pp}r^{pp}e^{2ik_{z0}d} = 0$. The dispersion curves are hyperbolic and unambiguously located in the bright contour of mode branches. Figure 5(b) shows the in-plane PTC for the shear-free biaxial material at $\omega = 90$ meV/$\hbar$, for which $\{\varepsilon_{xx}, \varepsilon_{yy}, \varepsilon_{zz}\} = \{-1.39 + 0.86i, 0.29 + 0.71i, -2.38 + 0.27i\}$. The topological structure of the hyperbolic polariton response at the given frequency is like that of the PTC in Fig. 5(a). However, for $\beta$-Ga$_2$O$_3$, the hyperbolic mode bright branches are rotated with respect to the coordinate system of the monoclinic plane. To describe the nature of this rotation, we show the rotated optical axes of this low-symmetry Bravais crystal by diagonalizing the real part of the permittivity tensor of $\beta$-Ga$_2$O$_3$. The rotation angle for optical axes is [48]

$$\gamma = \frac{1}{2}\mathrm{atan}\left[\frac{2\,\mathrm{Re}(\varepsilon_{xy})}{\mathrm{Re}(\varepsilon_{xx}) - \mathrm{Re}(\varepsilon_{yy})}\right]. \tag{6}$$

The rotated optical axes are shown in Figs. 5(d) and 5(e) (white dashed lines) and match the rotation of the shear hyperbolic polaritons. Moreover, when the off-diagonal component is introduced into the biaxial-hyperbolic material, shearing causes the hyperbolic mode resonance of the system to be confined to the narrower bright branches. This suppression of photon tunneling further explains the spectral recession resulting from shearing in Fig. 4. Meanwhile, the bright branches of the hyperbolic mode resonance in Figs. 5(d) and 5(e) break the symmetry with respect to the optical axes. This reduced symmetry in the PTC of $\beta$-Ga$_2$O$_3$ is a direct consequence of the lack of symmetry in its vibrational structure [48].

We now discuss how shear plays a role in photon tunneling supported by elliptical polaritons, as shown in Figs. 5(c) and 5(f). As shown in Fig. 2, when the frequency is 97 meV/$\hbar$, the polaritons of this low-symmetry Bravais crystal are in the elliptical mode. Figure 5(c) shows the in-plane PTC for the shear-free biaxial material system at $\omega = 97$ meV/$\hbar$. The PTC clearly shows an elliptical bright band, which is consistent with the results predicted by the permittivity (Re[$\varepsilon_{xx,yy,zz}$] < 0) in Fig. 2. To confirm the dominant role of elliptical polaritons in photon tunneling between shear-free biaxial materials, Fig. 5(c) shows the in-plane polariton resonance dispersion at the given frequency. The dispersion curves are elliptical and unambiguously located in the bright branches of elliptical mode resonance.

For a low-symmetry Bravais crystal system, the bright branches still have an elliptical topology [Fig. 5(f)], that is, the shear does not transform the topological structure of the elliptical polariton mode. The shear elliptical polaritons are tilted with respect to the coordinate system of the monoclinic plane, as anticipated by the optical axes [Fig. 5(f)], and this feature also appears clearly upon examining the in-plane polariton resonance dispersion. Moreover, since the propagation of polaritons is nontrivial in low-symmetry systems, the shear effect would intensify the in-plane anisotropy of the elliptical mode resonance. Photon tunneling in the low-symmetry Bravais crystal system has elliptical bright branches with a stronger anisotropy, as shown in Fig. 5(f). This conclusion is also supported by the in-plane polariton resonance dispersion of the elliptical mode resonances [Figs. 5(c) and 5(f)].

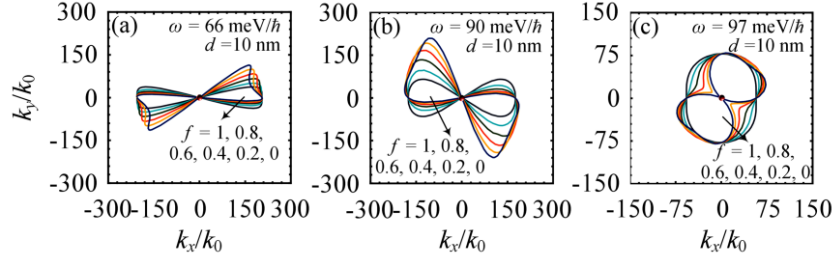

**Fig. 6.** Polariton resonance dispersions with different scaling factors at a frequency of (a) $\omega = 66$ meV/$\hbar$, (b) $\omega = 90$ meV/$\hbar$, and (c) $\omega = 97$ meV/$\hbar$.

Next, to better understand the mechanism behind these results, we analyze in Fig. 6 the polariton resonance dispersion for different scaling factors of the off-diagonal component. Figures 6(a) and 6(b) show that, when we remove the off-diagonal component ($f = 0$), the polariton resonance dispersion curves are hyperbolic and distributed symmetrically along the $x$ axis. Upon gradually increasing the magnitude of the off-diagonal terms, the polariton resonance dispersion curves become increasingly skewed with respect to the $x$ axis [Figs. 6(a) and 6(b)]. Moreover, as the scaling factor increases, the hyperbolic polariton resonance dispersion curves in Figs. 6(a) and 6(b) break the symmetry between the two sides of the hyperbola. For the elliptical polariton mode, the increase in scaling factor also deflects gradually the elliptical dispersion, as shown in Fig. 6(c). In addition, Fig. 6(c) shows how the off-diagonal terms affect the in-plane anisotropy of elliptical polariton modes. A higher scaling factor (a stronger shear effect) produces a stronger anisotropy in the elliptical dispersion.

The physical mechanism by which shear affects the twist-induced near-field thermal modulation is revealed in Fig. 7 by the photon tunneling feature of the low-symmetry Bravais crystal system with twist angle. To visually evaluate how shear affects the near-field thermal modulation, the photon tunneling feature of the higher-symmetry materials (shear-free biaxial material) is also shown in Fig. 7. These results demonstrate that appropriately tuning the twist angle significantly enhances the photon tunneling between the receiver and emitter. Let us first analyze the twist-induced feature of hyperbolic polaritons ($f = 0$) in Figs. 7(a) and 7(b). The resonance branches of shear-free biaxial material have a hyperbolic topology at a frequency of 67 meV/$\hbar$ [Fig. 7(a)], when the system is in the parallel case ($\theta = 0°$). Figure 7(b) shows that the increase in rotation angle remodels the bright branches of the PTC due to the twist-induced decoupling between the hyperbolic polariton modes of the receiver and emitter. The physical mechanism producing this twist-induced decoupling is the mismatch of the polaritons resonance in $k_x$-$k_y$ space. Notably, this decoupling phenomenon curtails the wavevector range of the coupling resonance and weakens the intensity of photon tunneling, which in turn attenuates the spectral HTC. As the scaling factor for the off-diagonal component returns to its natural value ($f = 1$), Fig. 7(c) shows that the hyperbolic polariton mode would be tilted by the shear effect and features symmetry breaking, as mentioned in the above discussion. This shearing effect, however, does not enhance the twist-induced decoupling between the hyperbolic polariton mode of the receiver and emitter. When the twist angle of the system is 90°, the PTC diagram of the system still has strong resonance branches over a broad range of wave vectors, deteriorating severely the twist-induced near-field thermal modulation supported by the shear hyperbolic polariton mode. Figure 7(i) shows the twist-induced modulated coefficient $\eta$ at the frequency of each polariton mode. The twist-induced modulated coefficient for the hyperbolic polaritons can reach 1.33, whereas the scenario for shear hyperbolic polaritons only produces a twist-induced modulated coefficient of 1.27, which is an attenuation of 4.51%.

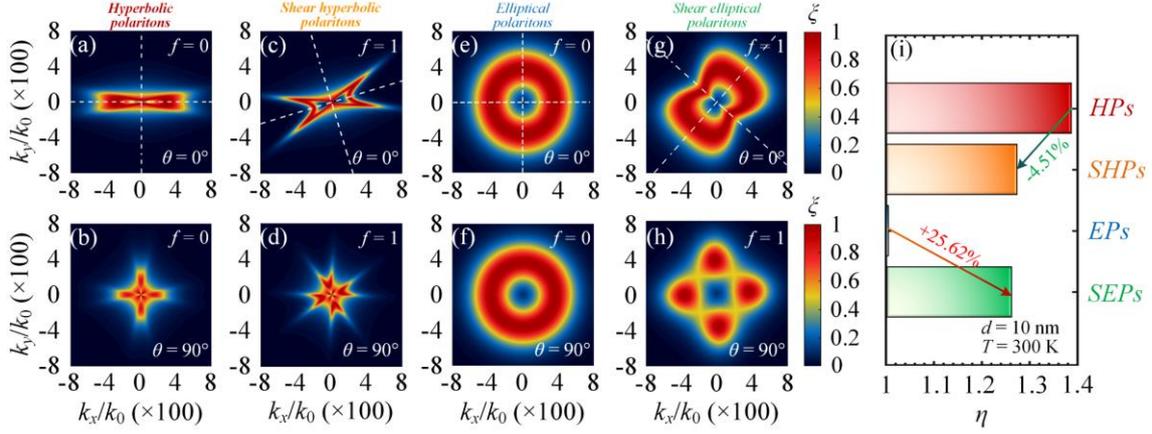

**Fig. 7.** PTCs for four types of surface modes at different twist angles. Four surface modes are considered: (a), (b) hyperbolic polaritons, (c), (d) shear hyperbolic polaritons, (e), (f) elliptical polaritons, and (g), (h) shear elliptical polaritons. (i) Twist-induced modulated coefficients for four types of surface modes. The white dashed lines indicate the optical axes.

The shear effect significantly improves the performance of twist-induced radiative modulation for the elliptical polariton mode. At a frequency of 94.5 meV/$\hbar$, the introduction of the off-diagonal component increases the modulated coefficient from 1.01 (elliptical polaritons) to 1.26 (shear elliptical polaritons), for an increase of 25.62% [see Fig. 7(i)]. This enhancement of radiative modulation can be elaborated in the photon tunneling feature of Figs. 7(e)–7(h). For the shear-free biaxial scenario, we notice that, as the emitter rotates, the weak intrinsic elliptical anisotropy [Fig. 7(e)] decouples the PTC [Fig. 7(f)], which is the main reason for the feeble twist-induced radiation modulation for the elliptical polariton mode at the given frequency. However, as mentioned above, the shear effect in a low-symmetry Bravais crystal intensifies the in-plane anisotropy of the elliptical polariton mode [Fig. 7(g)]. Notably, Fig. 7(h) shows that, as the twist angle increases, the region with high PTC gradually decreases, which limits the tunneling of thermal photons between the two $\beta$-Ga$_2$O$_3$ slabs. This is the main reason why the twist-induced modulated coefficient of the shear elliptical polariton mode is superior to that of the elliptical polariton mode in Fig. 7(i). Consequently, we conclude that the shear effect can be used to substantially improve the twist-induced near-field thermal modulation supported by the elliptical polariton mode, whereas the twist-induced thermal modulation is suppressed by the shear effect for the hyperbolic polariton mode.

## 4. CONCLUSION

To conclude, we have investigated radiative heat transfer between two low-symmetry Bravais crystal (*β*-Ga$_2$O$_3$) slabs. We demonstrate that *β*-Ga$_2$O$_3$ is a natural low-symmetry polaritonic material with a strong surface state, producing near four-order-of-magnitude enhancement of thermal radiation beyond the blackbody limit. It also significantly exceeds the NFTR from traditional dielectric materials, such as SiO$_2$, SiC, and *h*BN, that are widely used in the field of near-field radiative heat transfer.

In addition, we demonstrate that the intrinsic shear effect caused by twist-induced radiative modulation makes low-symmetry Bravais crystal systems superior to high-symmetry crystal systems. By introducing a scaling factor to describe the magnitude of the off-diagonal component, we visually show the evolution of surface polaritons in low-symmetry Bravais crystals exposed to the shear effect. Finally, we demonstrate that this effect results in twist-induced decoupling of hyperbolic and elliptical polaritons and substantially improves the twist-induced near-field thermal modulation supported by the elliptical polariton modes.

The results of this study show that low-symmetry Bravais crystals can serve as a prospective platform for high-performance radiative heat transport and twist-induced near-field thermal control. The phenomena demonstrated herein can also be extended to other low-symmetry materials, such as geological minerals, common oxides, and organic crystals. On the other hand, the reconstitution phenomenon on photon tunneling exposed to the shear effect may have important implications in the manipulation of phase and directional energy transfer (including ultra-fast thermal dissipation, thermomechanical stability, and radiative heat shuttling), which creates interesting opportunities for more efficient and powerful thermal photons computing and photonic energy-harvesting techniques.

**Funding**. National Natural Science Foundation of China (Grant No. 52076056), Fundamental Research Funds for the Central Universities (Grant No. FRFCU5710094020)

**Acknowledgment.** The authors thank A Paarmann for helpful discussion. C-L Zhou, Y Zhang, and H-L Yi are grateful to National Natural Science Foundation of China and the Fundamental Research Funds for the Central Universities.